\documentclass[aps,prl,twocolumn,amsmath,amssymb,floatfix]{revtex4}
\usepackage{tabularx}
\usepackage{bm}
\usepackage{euscript}
\usepackage{graphicx}
\usepackage{color}
\usepackage{amsfonts}
\usepackage{exscale}
\usepackage{amsbsy}

\begin{document}

\title{Detecting the Majorana fermion surface state of $^3$He-B through spin relaxation}


\author{Suk Bum Chung and Shou-Cheng Zhang}
\affiliation{Department of Physics, McCullough Building, Stanford University, Stanford, CA 94305}
\date{\today}

\begin{abstract}
The Majorana fermion, which can be useful for topological quantum
computation, has eluded detection. The 3He-B, recently shown to be a
time-reversal invariant topological superfluid, has a gapless Majorana
fermion surface state. We show here that an electron spin relaxation
experiment can detect this surface state - its Majorana nature through 
the Zeeman field direction dependence of the relaxation time 
$1/T_1 \propto \sin^2 \theta$, where $\theta$ is the angle between the 
field and the surface normal. We propose an experiment setup where an 
electron inside a nano-bubble is injected below the 3He liquid surface.
\end{abstract}

\maketitle

Recent development has secured for the Majorana fermion a central 
place in wide range of theoretical physics \cite{WILCZEK2009}. 
The chief characteristic of the Majorana fermion is that it has 
only {\it half} the degree of freedom as the usual complex fermions. 
It is due to this characteristic that if neutrions are Majorana fermions 
we can have neutrinoless double beta decay. In recent years, there 
has been great interest in condensed matter systems where Majorana 
fermions can arise. Systematic understanding of such systems has been 
obtained through investigating their topological properties, which 
were shown to be analogous to those of topological insulator (TI) 
\cite{BERNEVIG2006,KONIG2007,FU2007A,HSIEH2008,ZHANG2009,XIA2009,CHEN2009}.
Similar to TI, topological superconductors (SC) or superfluids
have a full pairing gap inside the bulk, but have protected gapless
state at the edge or on the surface. One example is the weak-pairing 
phase of two-dimensional (2D) spinless chiral SC with $p+ip$ symmetry 
\cite{volovik1988,volovik2003,READ2000,STONE2003}. This system breaks 
time reversal symmetry, and can be understood as the SC analogue of 
the quantum Hall (QH) state. The main difference is that the chiral 
edge state of the chiral SC consists of Majorana fermions rather 
than complex fermions as in the $\nu = 1$ QH state, and thus contains 
only half the degrees of freedom. In addition it was shown that a 
Majorana zero mode is trapped in each vortex core \cite{READ2000}, 
leading to the non-Abelian statistics of vortices \cite{IVANOV2001}. 
More recently the time-reversal invariant (TRI) SC has been proposed 
\cite{QI2009,roy2008} and classified \cite{SCHNYDER2008,KITAEV2009}. 
Such topological SC or superfluid states in two and three dimensions are 
the analogous to the TRI quantum spin Hall (QSH) or the TI state 
discovered recently \cite{QI2009,roy2008,SCHNYDER2008}. So  far, the 
only definite candidate for the 3D TRI topological SC state is the 
$^3$He-B phase \cite{QI2009,roy2008,SCHNYDER2008}, the topological 
invariant of which was first pointed out in Refs.~\cite{salomaa1988,volovik2003}. 
In fact, the Bogoliubov-de Genne (BdG) Hamiltonian for $^3$He-B phase is 
identical to the simplest model Hamiltonian of the 3D TI \cite{QI2008,ZHANG2009}, 
giving rise to a single surface state described by the Hamiltonian
\begin{equation}
\mathcal{H}_{surf} = v_F {\bm \sigma}\cdot({\bf \hat{z}}\times{\bf
p}),
\label{EQ:surfaceH}
\end{equation}
where ${\bf p}$ is the in-plane momentum, ${\bf \hat{z}}$ is the
surface normal, and ${\bm \sigma}$ is the dimensionless spin operator. 
Despite having the same Hamiltonian, the surface state of the 
$^3$He-B phase consists of a single Majorana cone which has only half 
the degrees of freedom as the surface state of the TI which consists 
of a single Dirac cone.

There has been recent experimental efforts to detect the surface 
states of $^3$He-B \cite{AOKI2005,CHOI2006}. Despite results 
consistent with the existence of the gapless Andreev bound state at the 
surface, 
these experiments were done on a `rough' surface and
did not directly detect the Majorana cone or the surface state 
degree of freedom being half that of the usual complex fermions. 
We need a probe for a free surface to detect the Majorana nature 
of the surface mode, i.e. an analogue of neutrinoless double beta 
decay. There are restrictions on external perturbations which can 
couple to the Majorana surface state of $^3$He-B; indeed as they are 
due to the halving of the degrees of freedom these restrictions are 
probably the most distinctive features of the surface state. The 
material properties of $^3$He-B, 
mainly its very low energy scale, impose further constraints on
possible experimental methods. Nonetheless, we find that the Majorana
nature of the surface mode gives rise to some striking and
qualitatively distinct experimental signatures.

{\it Surface state of Majorana fermion}: First, we show the basic 
similarity and difference between the surface modes of $^3$He-B and 
the simplest 3D TRI TI \cite{QI2008,ZHANG2009}. The $^3$He-B surface 
mode is derived from the BdG Hamiltonian,
\begin{equation}
\hat{\mathcal{H}}_{BdG} =\!\left[\begin{array}{cccc} \epsilon_{\bf p}\!-\!E_F & 0 & -\frac{\Delta}{p_F}\!\hat{p}_-  & \frac{\Delta}{p_F}\!\hat{p}_x\\
0 & \epsilon_{\bf p}\!-\!E_F & \frac{\Delta}{p_F}\!\hat{p}_x & \frac{\Delta}{p_F}\!\hat{p}_+\\
-\frac{\Delta}{p_F}\!\hat{p}_+ & \frac{\Delta}{p_F}\!\hat{p}_x & -\!\epsilon_{\bf p}\!+\!E_F & 0\\
\frac{\Delta}{p_F}\!\hat{p}_x & \frac{\Delta}{p_F}\!\hat{p}_- & 0 &  -\!\epsilon_{\bf p}\!+\!E_F \end{array}\right],
\label{EQ:BdG-BW}
\end{equation}
where we have used the basis
$\Psi_{BdG}({\bf r}) \equiv [\hat{\psi}_\rightarrow ({\bf r}),\hat{\psi}_\leftarrow ({\bf r}),
\hat{\psi}^\dagger_\rightarrow ({\bf r}), \hat{\psi}^\dagger_\leftarrow ({\bf r})]^T$
with the spin quantization axis along the $x$-axis (up to rotation
by the Leggett angle \cite{LEGGETT1975,vollhardt1990}
around the surface normal ${\bf \hat{z}}$).
$\epsilon_{\bf p} = p^2/2m$ is the free fermion Hamiltonian, $E_F$
is the $^3$He atom Fermi energy, and $\hat{p}_\pm = \hat{p}_y \pm i\hat{p}_z$.
As noticed in Ref. \cite{QI2009,roy2008}, $\hat{\mathcal{H}}_{BdG}$
is formally identical to the simplest model of TRI TI with the
surface state consisting of a single Dirac cone \cite{QI2008,ZHANG2009}.
In both cases, 
the momentum-dependence of the off-diagonal term 
leads to gapless modes bound to the surface \cite{BUCHHOLTZ1981,nagato1998}. 
The coupling of the spin and orbital degrees of freedom for the
surface state of the 3D TRI system can be understood simply by
setting $p_x =0$ in Eq.\eqref{EQ:BdG-BW} and reduce the system to
a 2D TRI system, described by the QSH model of Ref. \cite{BERNEVIG2006}.
This enables us to see that when the parallel momentum is aligned
along $y$-direction, the quasiparticle spin is polarized in
$\pm x$ direction and the $\rightarrow$($\leftarrow$)-spin surface
quasiparticle will have dispersion of $E = -(\Delta/k_F)k_y$
($E = (\Delta/k_F)k_y$). Due to invariance with respect to
simultaneous spin and orbital rotation around ${\bf \hat{z}}$,
this coupling of orbital and spin degrees of freedom holds for
all directions in the $xy$-plane.

Although the BdG Hamiltonian for $^3$He-B phase is formally 
similar to the model Hamiltonian for the simplest TI
\cite{QI2008,ZHANG2009}, the fermionic operators that form the bases 
of the two Hamiltonians are quite different. In $^3$He-B we have 
particle and hole excitations rather than conduction and valence 
band as in the TI. Since the spin-triplet pairing in $^3$He-B 
implies equal spin pairing, we cannot distinguish the particle and 
hole excitation through the spin degree of freedom, and thus the 
annihilation operator of the negative energy state is equivalent to 
the creation operator of the positive energy state. 

This Majorana nature of the $^3$He-B surface mode imposes strong
restriction on its interaction with an external perturbation. To see
how this restriction comes about, we need to examine the full mode
expansion of fermion creation and annihilation operators near the
surface. We impose the boundary conditions that the surface modes
vanish at the surface $z=0$ and decay exponentially in the $^3$He-B
liquid side (where $z<0$) of the surface, albeit much slower than
$k_F$. Since the wave vector parallel to surface ${\bf k}_\parallel$
remains a good quantum number, to satisfy these conditions the
surface modes needs to be proportional to
$e^{i{\bf k}_\parallel\cdot{\bf r}_\parallel}\sin(k_\perp z) e^{\kappa z}$,
where $k_F^2 = k_\parallel^2 + k_\perp^2$ and $\kappa > 0$. Inserting
this to the BdG equation Eq.\eqref{EQ:BdG-BW} gives
$\kappa = \Delta/\hbar v_F$ and reduces Eq.\eqref{EQ:BdG-BW} to an
effective surface Hamiltonian of Eq.\eqref{EQ:surfaceH}. Therefore,
for our surface mode expansion we use the result from the TI but also
take into account the artificial doubling mentioned above:
\begin{eqnarray}
\left[\begin{matrix} \hat{\psi}_\rightarrow({\bf r})\cr \hat{\psi}_\leftarrow({\bf r})\cr \hat{\psi}_\rightarrow^\dagger({\bf r})\cr \hat{\psi}_\leftarrow^\dagger({\bf r}) \end{matrix}\right] &=&  \sum_{\bf k} (\hat{\gamma}_{\bf k} e^{i{\bf k}_\parallel\cdot{\bf r}_\parallel} + \hat{\gamma}^\dagger_{\bf k} e^{-i{\bf k}_\parallel\cdot{\bf r}_\parallel})\left[\begin{matrix} \cos\frac{\phi_{\bf k}+\pi/2}{2} \cr \sin\frac{\phi_{\bf k}+\pi/2}{2} \cr \cos\frac{\phi_{\bf k}+\pi/2}{2} \cr \sin\frac{\phi_{\bf k}+\pi/2}{2} \end{matrix}\right]\nonumber\\
&\times& u_{\bf k}\!e^{\Delta z/\hbar v_F}\!\sin(k_\perp z)\!+\!({\rm gapped}\,\,\,{\rm modes}).
\label{EQ:modeMajorana}
\end{eqnarray}
where $\phi_{\bf k}=\arctan(k_y/k_x)$ and $u_{\bf k}$ is a
normalization constant of the mode ${\bf k}$ (see SOM for details).
Note that once we ignore the gapped modes (eigenenergy greater than
$\Delta$), we obtain the Majorana condition
$\hat{\psi}_\rightarrow({\bf r}) = \hat{\psi}_\rightarrow^\dagger({\bf r})$ and
$\hat{\psi}_\leftarrow({\bf r}) = \hat{\psi}_\leftarrow^\dagger({\bf r})$.
What this means is that the local creation and annihilation
operators for a fermion with its spin polarized parallel to the
surface is indistinguishable once we ignore modes with eigenenergy
greater than $\Delta$, thus reducing the degrees of freedom by 
half. Instead of the usual fermion anticommutation relation, these 
Majorana operators would form Clifford algebra,
$\sum_\sigma\{\hat{\psi}_\sigma({\bf r}), \hat{\psi}_{\sigma'}({\bf r'})\} = 2 \delta({\bf r}-{\bf r'})$
(where $\sigma,\sigma' = \rightarrow,\leftarrow$). It follows that
it is impossible to construct the spin-polarized local density
$\rho_{\sigma} ({\bf r})= \hat{\psi}_\sigma^\dagger({\bf r})\hat{\psi}_\sigma({\bf r})$
out of the gapless modes if the polarization axis is parallel to the
surface. This means that with the gapless surface mode, we can
neither construct the local density operator
$\rho ({\bf r}) = \sum_\sigma \hat{\psi}_\sigma^\dagger({\bf r})\hat{\psi}_\sigma({\bf r})$
nor the components of the local spin density operator parallel to
the surface,
$\hat{I}_x = (\hat{\psi}_\rightarrow^\dagger \hat{\psi}_\rightarrow -\hat{\psi}_\leftarrow^\dagger \hat{\psi}_\leftarrow)/2$ and
$\hat{I}_y = (\hat{\psi}_\rightarrow^\dagger \hat{\psi}_\leftarrow + \hat{\psi}_\leftarrow^\dagger \hat{\psi}_\rightarrow)/2$.
However, it is possible to construct the component of spin density
operator perpendicular to the surface,
$\hat{I}_z ({\bf r}) = -i\hat{\psi}_\rightarrow({\bf r})\hat{\psi}_\leftarrow({\bf r})$
\cite{STONE2003}. So in $^3$He-B the surface state does not
contribute to the local density fluctuation while its local spin
density is effectively {\it Ising} for $T \ll \Delta$, which
means that the local external perturbation can excite the surface
state only if it couples to $I_z$; this is a direct consequence of 
the halving of the degrees of freedom.

Therefore, to detect the surface state and its Majorana nature, it
is best to measure dynamic susceptibility arising out of these gapless
modes. 
From the discussion above we see that the dynamic spin 
susceptibility tensor of the surface state has only single nonzero 
component: $\chi^{zz}$, which we can calculate from 
Eq.\eqref{EQ:modeMajorana}. Anisotropy this drastic cannot be 
obtained from spin-orbit coupling of the complex fermions such as 
we see in the TI surface state. So we conclude that the resonant 
spin spectroscopy is the best probe for the Majorana surface mode. 
The extreme anisotropy of the spin susceptibility should be revealed 
through striking anisotropy in the spin spectroscopy. Due to the 
gapless dispersion, there will be no $e^{-\Delta/T}$ suppression 
of this anisotropy. We now need a spin probe that best fits the 
material property of $^3$He-B.

{\it ESR - spin spectroscopy}: We propose electron spin relaxation
(ESR) as the best spin spectroscopy on the $^3$He-B surface state.
Our basic idea is to introduce some extra electrons to $^3$He-B,
apply a weak DC magnetic field (which satisfies $H \ll T/\mu_B$;
note $\Delta/\mu_B \approx 26.2$G \cite{WHEATLEY1975}), excite the
electron spins through resonance, and then let these electron spins
relax through interaction with the surface state. This relaxation
process would probe the dynamic spin susceptibility of the $^3$He-B
in a way analogous to the way the nuclear magnetic relaxation (NMR)
is used to probe the dynamic spin susceptibility of electron in a
crystalline system. Such probe should reveal the drastic anisotropy
of the dynamic spin susceptibility of the surface state due to its
Majorana nature. More explicitly, we start from the spin relaxation
rate formula:
\begin{eqnarray}
\frac{1}{T_1} &=& \frac{T}{\hbar}\!\sum_{\bf q}\!\int\!\!dz_e\!\int\!\!dz\!\int\!\!dz'_e\!\int\!\!dz'  P(q,z_e)P(q,z'_e)\nonumber\\
&\times& A_+(\!{\bf q}\!,z\!-\!z_e\!) A_-(\!-{\bf q}\!,z'\!-\!z'_e\!)\!\frac{{\rm Im}\chi^{zz}(\!q,\!\omega_L\!; z,\!z'\!)}{\omega_L},
\label{EQ:spinRelax}
\end{eqnarray}
where $P(q,z_e)$ is the static form factor of the electron (obtained
from Fourier transforming the $xy$ coordinates of the probability
density of a single electron), $A_+$ is the component of the
interaction that flips the electron spin with respect to the
direction of the Zeeman field, $z(z')$ and $z_e(z'_e)$ are the
$z$-coordinates of the $^3$He atoms and the electron respectively,
and $\omega_L=g\mu_B/\hbar$ is the Larmor frequency of the electron.
This formula would look like the standard NMR relaxation formula
\cite{abragam1961} if we drop out the $z$ dependence, the electron
form factor $P$, and restore the isotropy of the dynamic spin
susceptibility. Eq.\eqref{EQ:spinRelax} implies the dependence of
$1/T_1$ on the direction of the Zeeman field, because $A_+$ couple 
$I_z$ to the component of the electron spin perpendicular to the 
Zeeman field. 

To illustrate this dependence on the Zeeman field direction, we
consider a simple contact interaction model for the coupling
between the electron and $^3$He atom spins. If we set the magnetic
field direction as ${\bf \hat{z'}} = {\bf \hat{z}}\cos \theta + {\bf
\hat{x}}\sin \theta$, we can write down the contact interaction as
$H_{contact} = -A_{contact} I_z S_z  = -A_{contact} I_z [S_{z'} \cos
\theta - \frac{1}{2}(S_+ + S_-)\sin\theta]$, giving us $A_+ =
A_{contact} \sin\theta$. Inserting this into
Eq.\eqref{EQ:spinRelax}, we obtain $1/T_1 \propto \sin^2 \theta$. In
other words, the electron spin does {\it not} relax at all for
perpendicular field! 
By contrast, the same model gives us $1/T_1$ independent of $\theta$
for the surface state of the simplest TI, 
${\bf q}$ summation canceling out the spin susceptibility anisotropy.

Realistic calculation can still give us this drastic anisotropy of
spin relaxation. In $^3$He-B, the main channel of spin-spin coupling
is the dipole-dipole interaction, mainly because an electron
strongly avoids contact with $^3$He atoms. 
With the dipole-dipole interaction, we do have coupling between
$I_z$ and $S_{x,y}$:
\begin{eqnarray}
H_D\!&=&\!\! -\frac{\mu_0}{4\pi}\frac{r^2{\bm \mu}_e \cdot {\bm \mu}_{{\rm He}} - 3({\bm \mu}_e \cdot{\bf r})({\bm \mu}_{{\rm He}} \cdot{\bf r})}{r^5}\nonumber\\
&=&\!\! -\!\frac{\mu_0 g \mu_B \gamma \hbar}{4\!\pi\!(\!r_\parallel^2\!+\!z^2\!)^{\frac{5}{2}}}I_z[(\!r_\parallel^2\!-\!2z^2\!)S_z - 3z\!(\!xS_x +\!yS_y)\!],
\label{EQ:dipole}
\end{eqnarray}
where $\gamma$ is the gyromagnetic ratio of a $^3$He atom and $g$ is
the Land$\acute{e}$ $g$-factor of an electron. However, for the electron
below the liquid surface, the $S_{x,y}$ terms of Eq.\eqref{EQ:dipole}
may have little effect; because $z>0$ for helium atoms `below' the 
electron and $z<0$ for helium atoms `above' the electrons, the coupling 
to $S_{x,y}$ from the helium atoms above cancels out the coupling to 
$S_{x,y}$ from the helium atoms below. Since the spin interaction is 
effectively Ising (that is, $H_D \propto -I_z S_z$), we have 
$1/T_1 \propto \sin^2 \theta$, as we argued the previous paragraph. 
By multiplying $\sin\theta$ to the 2D Fourier transform on the 
coefficient of the $I_z S_z$ term of Eq.\eqref{EQ:dipole}, we 
obtain $A_+({\bf q},z) = -\frac{\mu_0 g\mu_B \gamma\hbar}{2}q e^{-q|z|}\sin\theta$. 
As the next step, we need to devise an experimental setup to relax 
the electron spin by the $^3$He-B surface state. 

\begin{figure}
\begin{center}
\includegraphics[width=2.5in]{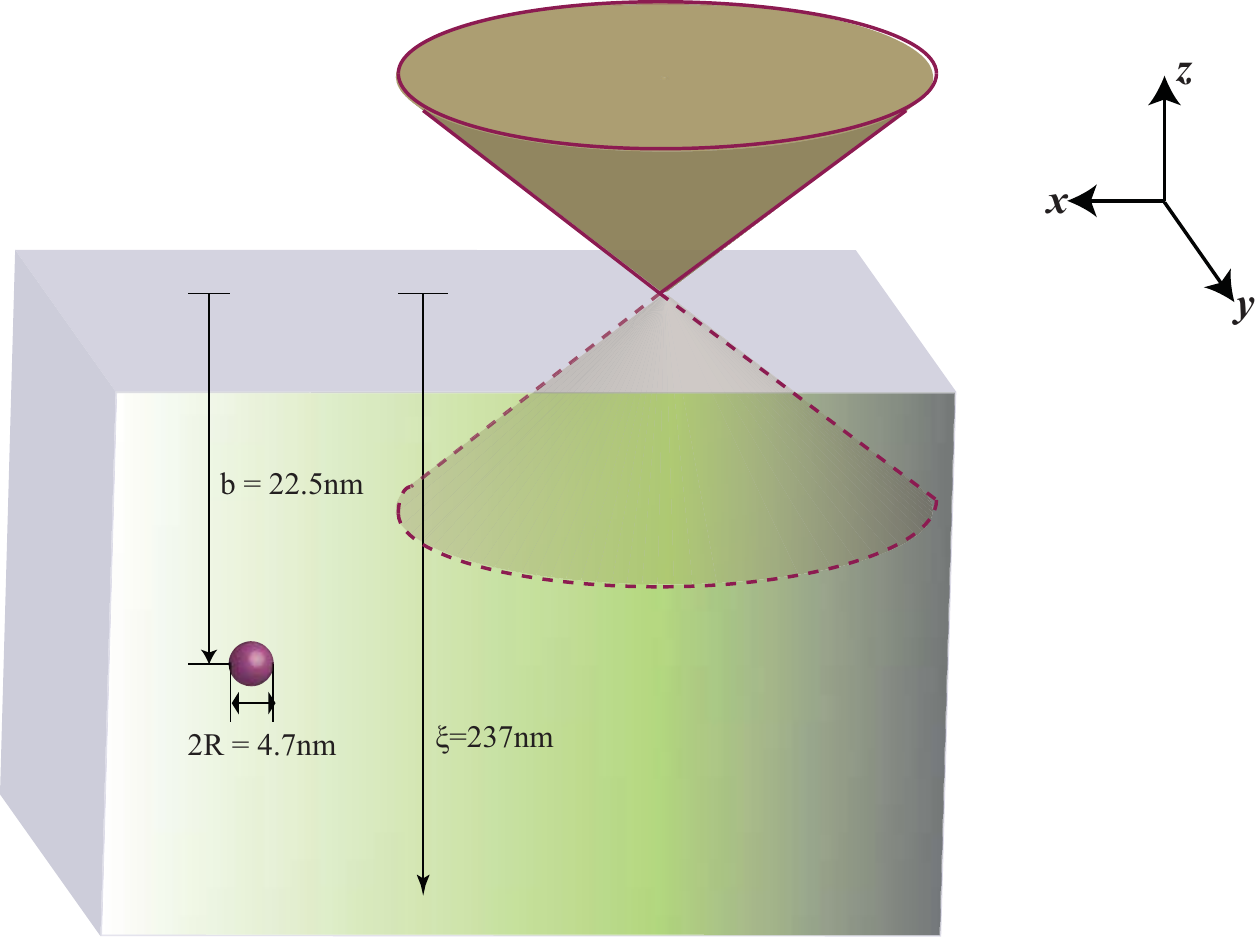}
\end{center}
\caption{Illustration of the surface state of $^3$He-B phase
consisting of a single Majorana cone, where the $E<0$ part of the
quasiparticle spectrum is redundant and indicated with the dashed
boundary. Also shown are the dimensions of the bubble electron when
we apply a perpendicular electric field of 150V/cm. Note how small
the size and depth of the bubble is compared to the depth $\xi$ of
the surface state, for which we take the weak coupling 
approximation $\hbar v_F/\Delta$ as in Eq.\eqref{EQ:modeMajorana}.}
\vspace{-6mm}
\label{FIG:bubble}
\end{figure}

{\it Electron bubble}: A crucial constraint on the relaxation rate
is how well the electron is localized. 
Whereas in the NMR, we can assume that a nucleus is a point-like
object, we cannot make the same assumption for electrons in ESR and
hence the introduction of the static form factor $P(q)$ in
Eq.\eqref{EQ:spinRelax}. Due to the Heisenberg uncertainty
principle, the more delocalized the electron is in the real space,
the more rapidly $P(q)$ falls off with $q$. This suppresses the
spin relaxation for processes that result in a large momentum change
for $^3$He atoms 
and hence suppresses $1/T_1$. For this reason, $1/T_1$ is very small
for an electron sitting on top of the $^3$He liquid surface. 
Even when electrons above the surface form a Wigner crystal, the
zero-point displacement is greater than 10\% of the lattice constant
for the lattice constant $\lesssim 1\mu$m \cite{monarkha2004}. There
is a limit to reducing the lattice constant as we need to keep the
dipole-dipole interaction between adjacent electrons much weaker than
the interaction between an electron and $^3$He atoms. In order to
enhance the electron localization significantly, we need to place the
electron under the $^3$He liquid surface. 

Once it is injected below the $^3$He liquid, an electron settles into
a well-localized metastable state below the surface. It cannot be easily
ejected from the liquid due to an electrostatic energy barrier at the
surface arising from the fact that helium is dielectric
\cite{COLE1974}. 
From the electric field boundary condition
$E_z|_{z=0+} = \epsilon_d E_z|_{z=0-}$ we see that the polarization of
atoms causes the surface of dielectric to repel an electron below the
surface. By tuning an electric field perpendicular to the surface, we can
adjust the equilibrium distance $|b|$ between the electron and the liquid
surface from right near the surface, $\sim$10nm, to below the surface
state, $|b|>\xi$ \cite{COLE1974,SHIINO2003}. Below the liquid surface, an
electron opens up a nano-sized cavity and becomes trapped inside of it
to avoid the energy cost due to the negative electron affinity of helium
atoms. The size of this `bubble' is determined by competition between the
zero-point kinetic energy of the confined electron $E_{ZP} = h^2/(8mR^2)$
and the surface energy of the cavity $E_S = 4\pi R^2 \alpha$, where $R$
is the cavity radius and $\alpha=$0.156 erg/cm$^2$ is the surface tension
of the helium liquid \cite{SUZUKI1988}. This gives us the electron
localization $R = [h^2/(32\pi m \alpha)]^{\frac{1}{4}}=$2.35nm, far better
than what we obtain above the surface. Fig.~1 shows this electron bubble
radius compared to the depth of the surface state.

\begin{figure}
\begin{center}
\includegraphics[width=.31\textwidth]{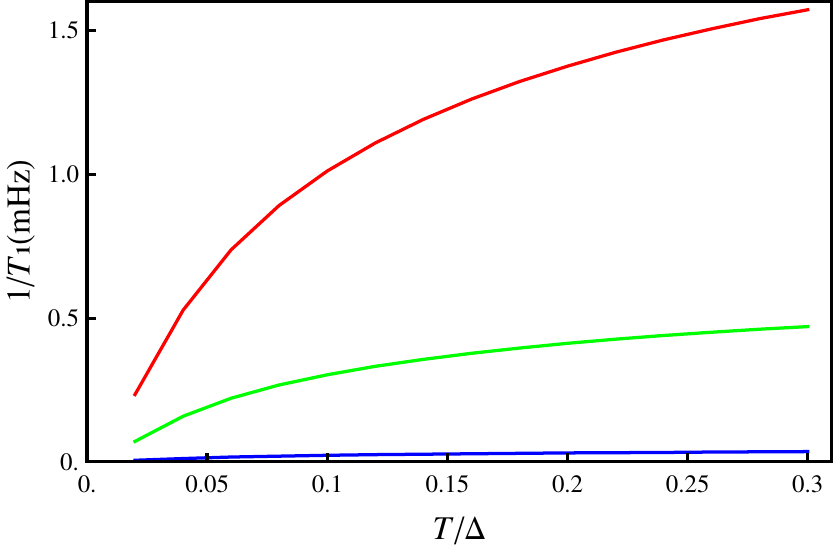}
\end{center}
\caption{The electron spin relaxation rate (in $10^{-3}$Hz) due to 
the surface state through dipole-dipole interaction for the 
magnetic field applied parallel to the surface. From the top to the 
bottom curve, the applied perpendicular electric field is 150V/cm, 
10V/cm, and 1.5V/cm respectively, which gives us the bubble depth 
$|b|$= 22.5nm, 87.4nm, and 225.2nm respectively 
(for consistency with Eq.\eqref{EQ:modeMajorana} the depth of the 
surface state is set $\xi = \hbar v_F/\Delta 237$nm ). 
}
\vspace{-6mm}
\label{FIG:relax}
\end{figure}

Our ESR rate calculation shows signatures of both the Majorana nature
and gapless dispersion. For electron bubbles placed at 22.5nm, 87.4nm,
and 225.2nm below the surface, we find that the relaxation rate is
$5 \times 10^2$ times faster for the parallel field ($\theta = \pi/2$)
than for the perpendicular field ($\theta = 0$), implying that we
effectively have $1/T_1 \propto \sin^2 \theta$ relation. As shown in
Fig.~2, for the bubble depth of 22.5nm, the relaxation rate $1/T_1$ is
approximately $10^3$ sec (see SOM for details) The absence of the
$\exp[-\Delta/T]$ suppression in $1/T_1$ versus $T$ behavior
characteristic of the bulk quasiparticle is the consequence of the 
gapless dispersion on the surface. However the relaxation rate 
anisotropy will be reduced if we include contribution from bulk 
condensate, which has isotropic nonzero spin susceptibility 
\cite{LEGGETT1975,vollhardt1990}.

In conclusion we have proposed a realistic experiment setup to
observe the Majorana fermion surface states of the topological
superfluid $^3$He-B phase. Due to the Majorana nature of the 
surface state, the spin density operator is purely Ising-like,
polarized perpendicular to the surface. Through an ESR experiment, 
we can show both gapless dispersion and extreme anisotropy of the 
dynamic spin susceptibility. Our experimental setups for the ESR 
measurement uses electron nano-bubbles placed below the liquid 
helium surface, giving rise to the $1/T_1 \propto \sin^2\theta$ 
dependence on the magnetic field direction. Such a direct 
experimental observation of the Majorana fermion would enhance our 
fundamental understanding this exotic particle and the nature of the 
topological superfluid, and pave the way for topological quantum 
computing.

{\it Acknowledgement}: We owe special thanks to K. Kono for
teaching us the electron bubble formation in helium liquid and X.-L. 
Qi for help in calculating dipole interaction in the momentum space. 
We also would like to thank M. Stone, A. Fetter, W. Halperin, 
D. Scalapino, D. Osheroff, S. Raghu, T. Hughes, and J. Maciejko for 
insightful discussions. This work is supported by DOE under contract 
DE-AC02-76SF00515 and Stanford ITP.

{\it Notes added}: Near the completion of this work, we learned 
that Nagato {\it et al.} independently obtained the spin 
susceptibility anisotropy of the surface state \cite{nagato2009}.

\vspace{-5mm}


\appendix
\section{The supporting online material}
\subsection{Surface state mode expansion}

In solving the BdG Hamiltonian Eq.(2), we make the following weak-pairing 
approximation:
\begin{eqnarray}
\epsilon_{\bf p}-E_F &=& -i\frac{\hbar^2}{m}\sqrt{k_F^2 - k_\parallel^2} \partial_z,\nonumber\\
\frac{\Delta}{p_F} \hat{p}_\pm &=& \Delta \left[\frac{k_\parallel}{k_F}\sin\phi_{\bf k} \pm i \sqrt{1 - (k_\parallel/k_F)^2}\right],\nonumber\\
\frac{\Delta}{p_F} \hat{p}_x &=& \Delta \frac{k_\parallel}{k_F}\cos\phi_{\bf k}.
\end{eqnarray}
Now when the in-plane momentum is aligned along $y$-direction, other than
the quasiparticle spin-polarization in $\pm x'$ direction, it is
essentially identical to the 2D SC with $p_y+ip_z$ pairing for
$|\rightarrow\rightarrow\rangle$ pairs and $p_y-ip_z$ pairing for
$|\leftarrow\leftarrow\rangle$ pairs. From the edge state of such SC, we
can see that with the in-plane momentum aligned along $y$-direction,
$\rightarrow$($\leftarrow$)-spin surface quasiparticle will have dispersion
of $E = -(\Delta/k_F)k_y$ ($E = (\Delta/k_F)k_y$). Such alignment of orbital
and spin degrees of freedom should hold for all direction in the $xy$-plane.
Therefore the full mode expansion gives us Eq.(3). Note that neither 
introducing anisotropy between the in-plane ($\hat{p}_\pm \Delta/p_F$) and 
perpendicular ($\hat{p}_z \Delta/p_F$) components of the gap nor taking into 
account the possible $z$ dependence of the perpendicular component of the 
gap is going to change the mode expansion qualitatively.

\subsection{Calculating the relaxation rate}

To calculate the spin relaxation rate Eq.(4) due to the
dipole-dipole interaction of Eq.(5), we need to calculate from $H_D$ 
the matrix element that couples to $S_-=(S_x \cos\theta - S_z \sin\theta)-iS_y$ 
in the momentum space. To obtain this, we define ${\bf A}_D ({\bf r})$ from 
$H_D \equiv I_z({\bf r}) {\bf A}_D ({\bf r}-{\bf r}_e) \cdot {\bf S}({\bf r}_e)$. 
2D Fourier transform gives us
\begin{equation}
{\bf A}_D ({\bf q},z) = \frac{\mu_0 g\mu_B \gamma\hbar}{2} e^{-q|z|} (iq_x {\rm sgn}(z), iq_y {\rm sgn}(z), q).
\label{EQ:dipoleQ}
\end{equation}
From this, we can see that the component of ${\bf A}_D$ that couples to 
$S_-$ is 
\begin{eqnarray}
A_D^+ ({\bf q},z) &=& A_D^x\cos\theta +iA_D^y - A_D^z\sin\theta\nonumber\\
&=& \frac{\mu_0 g\mu_B \gamma\hbar}{2} e^{-q|z|}[(iq_x\cos\theta - q_y){\rm sgn}(z)-q\sin\theta].\nonumber\\
\label{EQ:dipoleQ2}
\end{eqnarray}

In calculating the imaginary part of the dynamic susceptibility,
we take the $\omega_L \to 0$ limit. In this limit, the anomalous
part of the Green function do not contribute. This means that
$\chi^{zz}$ is the same as the 3D strong topological insulator
surface state with the chemical potential at the Dirac point.
Therefore, the imaginary part of the dynamic spin susceptibility
in this limit is
\begin{eqnarray}
&&\frac{{\rm Im}\chi^{zz}({\bf q},\omega_L;z,z')}{\omega_L}\nonumber\\
&=&\pi e^{2(z+z')/\xi}\int \frac{d^2 k}{(2\pi)^2}\left(-\frac{\partial f}{\partial E_{\bf k}}\right)\delta(E_{{\bf k}+{\bf q}}-E_{\bf k})\nonumber\\
&\times& u^2_{{\bf k}+{\bf q}} u^2_{\bf k}\sin^2\frac{\phi_{{\bf k}+{\bf q}}-\phi_{\bf k}}{2}\nonumber\\
&\times&\sin(\sqrt{k^2_F - k^2}z)\sin(\sqrt{k^2_F - ({\bf k}+{\bf q})^2}z)\nonumber\\
&\times&\sin(\sqrt{k^2_F - k^2}z')\sin(\sqrt{k^2_F - ({\bf k}+{\bf q})^2}z')\nonumber\\
&=&\frac{k^2_F}{4\pi\Delta^2}e^{2(z+z')/\xi}\int^{k_F}_{q/2}dk u^4_{\bf k}\left(-\frac{\partial f}{\partial k}\right)\frac{q/2k}{\sqrt{1-(q/2k)^2}}\nonumber\\
&\times&\sin^2(\sqrt{k^2_F - k^2}z)\sin^2(\sqrt{k^2_F - k^2}z').
\label{EQ:suscep}
\end{eqnarray}

Lastly, there is the static form factor of the electron $P(q,z)$.
This is a 2D Fourier transform of the modulus square of the single
electron ground state wave function. In the case of the electron in a 
bubble, we can approximate the electron wave function to vanish at the 
bubble boundary, so we can set for the modulus square of the ground 
state wave function
\begin{equation}
P(r) = \frac{1}{2\pi R}\frac{\sin^2 (\pi r/R)}{r^2},
\end{equation}
where $R$ is the bubble radius, if we take the center of the bubble to
be the origin. A {\it 3D} Fourier transform of $P(r)$ approximates to
$\exp[-(R^2/2\pi^2)(q^2 + q^2_\perp)]$, so we can make approximation
\begin{equation}
P(q,z) \approx \frac{\sqrt{2\pi}}{R}e^{-q^2 R^2/2\pi^2}e^{-\pi^2 (z-b)^2/2R^2},
\label{EQ:form}
\end{equation}
where we now take the coordinate of the center of the bubble to be
$(0, 0, b)$. We also note that since the energy difference between the
electronic ground state to the first excited states is much larger than
the pairing gap of $^3$He-B.

With a given bubble depth $|b|$, we can calculate the ESR relaxation
rate from Eqs.\eqref{EQ:dipoleQ2}, \eqref{EQ:suscep}, and
\eqref{EQ:form}. We find that the relaxation rate due to the helium
atoms in the $|z-b|<R$ region is only about 4\% of the contribution
from the $|z-b|>R$ region. Therefore we only give the result for the
latter region. The result can be expressed in most part by dimensionless
integrals. For the Zeeman field parallel to the surface ($\theta = \pi/2$)
we obtain
\begin{eqnarray}
\frac{1}{T_{1\parallel}^{(1)}} &=& \frac{1}{8\pi^2}\frac{\Delta}{\hbar}\left(\frac{\mu_0 g \mu_B \gamma \hbar k_F^3}{2\Delta}\right)^2 e^{-4b/\xi}\nonumber\\
&\times& \int^1_0 dx \frac{x^4 e^{x/\tilde{T}}}{(1+e^{x/\tilde{T}})^2}\int^1_0 dy \frac{y^4}{\sqrt{1-y^2}}\nonumber\\
&\times&[{\rm erf}(\pi/\sqrt{2}-\sqrt{2}xyk_F R/\pi)\nonumber\\&+&{\rm erf}(\pi/\sqrt{2}+\sqrt{2}xyk_F R/\pi)]^2\nonumber\\
&\times&\!\!\left[\frac{e^{\frac{2R}{\xi}}-e^{\frac{2b}{\xi}}e^{-2xy k_F (b-R)}}{xy k_F \xi - 1}\!+\!\frac{e^{-\frac{2R}{\xi}}}{xy k_F \xi + 1}\right]^2,\nonumber\\
\label{EQ:relaxPar}
\end{eqnarray}
where $x = k/k_F$, $y = q/2k$ and $\tilde{T} = k_B T/\Delta$. For the
perpendicular ($\theta = 0$) Zeeman field, we obtain the relaxation rate
\begin{eqnarray}
\frac{1}{T_{1\perp}^{(1)}} &=& \frac{1}{8\pi^2}\frac{\Delta}{\hbar}\left(\frac{\mu_0 g \mu_B \gamma \hbar k_F^3}{2\Delta}\right)^2 e^{-4b/\xi}\nonumber\\
&\times& \int^1_0 dx \frac{x^4 e^{x/\tilde{T}}}{(1+e^{x/\tilde{T}})^2}\int^1_0 dy \frac{y^4}{\sqrt{1-y^2}}\nonumber\\
&\times&[{\rm erf}(\pi/\sqrt{2}-\sqrt{2}xyk_F R/\pi)\nonumber\\&-&{\rm erf}(\pi/\sqrt{2}+\sqrt{2}xyk_F R/\pi)]^2\nonumber\\
&\times&\!\!\left[\frac{e^{\frac{2R}{\xi}}-e^{\frac{2b}{\xi}}e^{-2xy k_F (b-R)}}{xy k_F \xi - 1}\!+\!\frac{e^{-\frac{2R}{\xi}}}{xy k_F \xi + 1}\right]^2.\nonumber\\
\label{EQ:relaxPerp}
\end{eqnarray}
We see in Eqs.\eqref{EQ:relaxPar} and \eqref{EQ:relaxPerp} that 
${\rm sgn}(z)$ in Eq.\eqref{EQ:dipoleQ2} leads to the cancelation
between contributions from $-b+R<z<0$ and from $z<-b-R$. For
$k_F = 7.88/$nm, $\xi = 237$nm, $R = 2.35$nm, Eq.\eqref{EQ:relaxPar}
is larger than Eq.\eqref{EQ:relaxPerp} by thee orders of magnitude
at $|b| =$ 22.5nm, 87.4nm, and 225.2nm; the dependence on $|b|$ is
quite weak. Fig.~2 is the plotting of Eq.~\eqref{EQ:relaxPar} for
different values of $|b|$.

Finally, we note that the relation between the applied perpendicular
electric field and the bubble depth is
\begin{equation}
|b_{equil}| = \left(\frac{e}{4\pi\varepsilon_0 E} \frac{\varepsilon_d - 1}{4\varepsilon_d(\varepsilon_d+1)}\right)^{1/2}.
\end{equation}
This comes from noting that a charge $q$ below dielectric (ratio
$\varepsilon_d$) surface at this depth ($z=b$; note $b<0$) induce
charge on dielectric surface through polarization. This surface charge
is effectively equivalent to having an image charge
$q(\varepsilon_d - 1)/(\varepsilon_d+1)$ at $z=-b$, which means that
the dielectric surface repels the charge below the surface. Therefore,
when we apply the constant perpendicular electric field $E$, the net
force the electron bubble feels is
\begin{equation}
F_z(b) = -\frac{e^2}{4\pi\varepsilon_0 (2b)^2}\frac{\varepsilon_d - 1}{\varepsilon_d(\varepsilon_d+1)} + eE.
\end{equation}
For helium-3 liquid $\epsilon_d = 1.04$ \cite{kierstead1976}; this 
gives $|b_{equil}|$ =22.5nm for $E =$ 150V/cm. We have simply set 
$b=b_{equil}$ and have not considered any fluctuation around this 
equilibrium point.

\end{document}